# Temperature dependence of Coulomb oscillations in a few-layer two-dimensional WS$_2$ quantum dot

Xiang-Xiang Song,[1,2] Zhuo-Zhi Zhang,[1,2] Jie You,[1,2] Di Liu,[1,2] Hai-Ou Li,[1,2] Gang Cao,[1,2] Ming Xiao,[1,2] and Guo-Ping Guo[1,2]

[1]*Key Laboratory of Quantum Information, CAS, University of Science and Technology of China, Hefei, Anhui 230026, China*

[2]*Synergetic Innovation Center of Quantum Information & Quantum Physics, University of Science and Technology of China, Hefei, Anhui 230026, China*

Standard semiconductor fabrication techniques are used to fabricate a quantum dot (QD) made of WS$_2$, where Coulomb oscillations were found. The full-width-at-half-maximum of the Coulomb peaks increases linearly with temperature while the height of the peaks remains almost independent of temperature, which is consistent with standard semiconductor QD theory. Unlike graphene etched QDs, where Coulomb peaks belonging to the same QD can have different temperature dependences, these results indicate the absence of the disordered confining potential. This difference in the potential-forming mechanism between graphene etched QDs and WS$_2$ QDs may be the reason for the larger potential fluctuation found in graphene QDs.

*Correspondence and requests for materials should be addressed to G.P.G. (gpguo@ustc.edu.cn).

After demonstrating a variety of electronic applications on graphene,[1] graphene-like two-dimensional layered materials have attracted the increasing attention of researchers because of their unique properties.[2,3] Compared with traditional semiconducting materials, two-dimensional layered materials have the advantages of inherent flexibility and an atomically-thin geometry. Moreover, because their interfaces are free of dangling bonds, two-dimensional layered materials can easily be integrated with various substrates. They can also be fabricated in complex sandwiched structures[4] or even suspended to avoid the influence of the substrate.[5]

Transition metal dichalcogenides (TMDCs) are considered one of the most promising two-dimensional post-graphene materials for electronics owing to the presence of a band gap ranging from 1 to 2 eV.[6,7] This band gap results in high on–off ratios in classical electronic devices made out of TMDCs.[8] A variety of electronic devices have been demonstrated on TMDCs, such as field-effect transistors,[9-13] heterostructure junctions[14,15] and photodetectors.[16,17]

Meanwhile, from the aspect of quantum nano-devices, TMDCs have advantages over graphene as well. Unlike etched quantum dots (QDs) made on semimetallic graphene,[18,19] semiconducting TMDCs have a band gap large enough to form a QD using the electric field. This QD decreases the influence of the edge states, which are a major aspect limiting the performance of graphene nano-devices.[20,21] Recently, theoretical predictions using QDs on TMDCs as qubits have been proposed,[22] and experimental studies of QDs on $WSe_2$ have been performed as well.[23] However, a detailed study of QDs in TMDCs, and especially the behavior of the Coulomb oscillations, is still not yet available.

Tungsten disulfide (WS$_2$) is a typical semiconducting TMDC material that has a direct band gap of 2.0 eV in a monolayer and an indirect band gap of 1.4 eV in the bulk crystal.[6] Theoretical models predict that WS$_2$ should have the highest mobility among the semiconducting TMDCs because of the reduced effective mass.[24] Several previous studies have shown that the typical carrier mobility of WS$_2$ is in the range of 20–300 cm$^2$/(V s).[11,25,26]

In this letter, we use standard semiconductor fabrication techniques to fabricate a QD made of WS$_2$, where Coulomb oscillations are observed. The temperature dependence of the Coulomb peaks is investigated, showing that the behavior of Coulomb peaks is in agreement with standard semiconductor QD theory: the full-width-at-half-maximum (FWHM) of the Coulomb peaks increase linearly while the peak height remain almost unchanged when increasing the temperature. The Coulomb oscillations of the QD on WS$_2$ are different from those of the graphene etched QDs, wherein Coulomb peaks belonging to the same QD can have different temperature dependences. Among these Coulomb peaks, only portions can be understood by the standard semiconductor QD theory, while most of the Coulomb peaks of graphene etched QDs become broader and higher with increasing temperature. Our results indicate the absence of the disordered confining potential, showing a difference between the potential-forming mechanisms of graphene etched QDs and WS$_2$ QDs, which may be the reason for the larger potential fluctuation found in graphene QDs. Meanwhile, the results may be useful for the fabrication of QDs on graphene-like two-dimensional layered materials.

## Results

As shown in Fig. 1(a), the few-layer $WS_2$ flake used in the experiment has an area of approximately $3\times5$ μm$^2$, which is highlighted by the white dotted line in the scanning electron microscope image. A schematic depiction of the device is shown in Fig. 1(b). The heavily doped silicon substrate worked as the back gate, which was isolated from the $WS_2$ by a 100 nm-thick layer of $SiO_2$. Here, we used Pd/Au for source-drain contacts ("source" and "drain" in Fig. 1(b)) of the $WS_2$ flake. On the $Al_2O_3$ insulating layer, four split electrodes ("top gates" in Fig. 1(b)) were fabricated using Ti and Au, corresponding to Middle Gate (MG), Left Barrier (LB), Plunger Gate (PG) and Right Barrier (RB) in Fig. 1(a). The voltage applied to MG, LB, and RB is mainly to form the confining potential to obtain quantum dot. The effect of PG is not only for confining the dot, but also for tuning the energy levels in the quantum dot.

Fig. 1(c) shows the schematic cross-section of the device. The experiment was performed in a He3 refrigerator at a base temperature of 240 mK. We used the standard lock-in method to probe the electronic signals.

First, we applied a dc voltage to the back gate to accumulate carriers in the $WS_2$ device. The source-drain current is obtained while sweeping the back gate voltage $V_{BG}$. As shown in Fig. 2(a), the characteristic behavior of an n-doped semiconductor is observed, that when tuning the back gate voltage to be more positive, the current becomes nonzero around $V_{BG}$=5 V. Note that the puddles may be formed in our device since some current oscillations are already evident. After we apply a dc voltage of −2 V to the top gates, the current as a function of the dc bias voltage at varying back gate voltages is measured. As shown in Fig. 2(b), over 20 consecutive Coulomb diamonds are observed from $V_{BG}$=13.2–14 V, indicating the formation of a QD. The fact that the Coulomb diamonds are symmetric suggests equivalent tunnel coupling to the source and drain leads. All of the Coulomb diamonds appear to have a similar charging energy $E_C$ of approximately 0.75 meV in logarithmic plotting (Fig. S1 in the supplemental material).

We also investigate the peak spacing of the Coulomb peaks. Fig. 2(c) shows the relative peak position as a function of peak number p, for the first 10 peaks of Fig. 2(b). The relative peak position can be expressed as V(p)−$V_0$, where V(p) is the voltage of the peak p and $V_0$ is the voltage of the first peak. The data was fitted by a linear function, as shown by red dashed line, suggesting a constant peak spacing of 36.6 mV. Using the $E_C$ and peak spacing $\Delta V$, the lever arm of the back gate $\alpha_{BG}$ is calculated to be 0.0205 eV/V. It is also estimated that the average QD radius is 430 nm using the relationship $E_C=e^2/(8\varepsilon_0\varepsilon_r r)$,[27] where $\varepsilon_r$=7 is the relative permittivity of few-layer $WS_2$ [32] and $r$ is the radius of the QD. Note that because the estimated dot size is comparable to the thickness of the insulating $SiO_2$ layer, the isolated disk approximation should change to a parallel plate approximation. Thus, the result of the estimated QD size should be considered as an upper limit, which is comparable to our geometric design of the top gates.

Further, to confirm that the QD is formed because of the presence of top gates, we use a plunger gate (PG) to tune the energy levels in the QD. As shown in Fig. 2(d), a Coulomb diamond is observed. These results are similar to those obtained in previous experiments on WSe$_2$.[23] There are two reasons why we don't have a complete Coulomb diamond in Fig. 2(d): (1) The thickness of the insulating layer Al$_2$O$_3$ is as larger as 100 nm. This will result in a relatively small lever arm α of the top gate, since α~ε/d, where ε is the relative dielectric constant of the insulating layer, and d is the thickness of the insulating layer. (2) Due to the limitation of ALD-growth quality, ε may also be smaller than theoretical expectation, resulting in a smaller lever arm as well. Moreover, due to the growth quality limitation, applying high voltage to top gates may lead to current leakage in the insulating layer (as shown in Fig. 2(d)). We only applied the voltage to show a dependence on $V_{PG}$. In fact, comparing the resonant tunneling point on Fig. 2(b) and Fig. 2(d), we found a current difference of about ~10 pA, which means there may be leakage between the source and the drain. We think the leakage is caused by the insulating layer of Al$_2$O$_3$. Using the leak current ~10 pA, divided by the voltage applied on the top gates, we can estimate that the resistance of the Al$_2$O$_3$ layer is about $5\times10^{11}\Omega$. Although the resistance is large, it may cause a leakage of several pA, which can be measured in the experiment.

Next, we investigate the temperature dependence of the Coulomb peak to achieve a better understanding of the Coulomb oscillations in WS$_2$. According to the standard theory of semiconductor QDs,[28] the line shape of the Coulomb peak in the weak coupling regime manifests in different forms for different temperature regimes, such that

$$G/G_{max} = \frac{1}{2}\cosh^{-2}(\frac{\delta}{2.5k_BT}) \quad \text{if } h\Gamma, \Delta E \ll k_BT \ll e^2/C, \quad \text{(classical regime)} \quad (1)$$

$$G/G_{max} = \frac{\Delta E}{4k_BT}\cosh^{-2}(\frac{\delta}{2k_BT}) \quad \text{if } h\Gamma \ll k_BT \ll \Delta E, e^2/C, \text{(quantum regime)} \quad (2)$$

where $G$ is the conductance of the Coulomb peak, $G_{max}$ is the maximum conductance at high temperature, $\Gamma$ is the tunneling rate through the barriers connecting the QD and the reservoirs, $\delta$ is the distance to the peak center in terms of energy, $\Delta E$ is the average level spacing, and $e^2/C$ is the charging energy. If the QD is in the strong coupling regime, the formula that describes the line shape of the peak changes to the well-known Breit–Wigner formula,

$$G = \frac{2e^2}{h}\frac{(h\Gamma)^2}{(h\Gamma)^2+\delta^2} \quad \text{if } T = 0, e^2/C \ll h\Gamma, \Delta E. \quad (3)$$

The Coulomb peaks in the regime $k_BT \sim h\Gamma$ have the Lorentzian lineshape of Eq. (3), where $k_B$ is Boltzmann's constant and $T$ is the temperature.[27]

Fig. 3(a) shows a typical Coulomb peak from the QD on WS$_2$, where the data (blue open squares) are fitted using both a Lorentzian curve (red solid line) and a cosh($x$) curve (blue solid line). The peak is also plotted in the logarithmic scale (Fig. 3(a) inset). The data decrease linearly at the tails of the peak, showing a deviation from the Lorentzian line shape. Note that only when $k_BT \geq h\Gamma$ does the slower decay of the Lorentzian tails become clearly visible.[27,29] According to the fitting curve, it is estimated that $h\Gamma$ of the QD is not dominated compared to the temperature.

To give a rough estimation of the order of magnitude, we can estimate the average single particle energy of the QD to be $\Delta E$=1.4 µeV using $\Delta E=\hbar^2/m^*r^2$,[27] where $r$ is the radius of the QD and $m^*$=0.3$m_e$ is the effective mass in WS$_2$.[22] Because the temperature varies in the range 290 mK to 1K, it can be estimated that $k_BT$~50 µeV, which suggests that the QD is in the classical regime ($\Delta E \ll k_BT$). We fit the Coulomb peak at different temperatures using Eq. (1). Fig. 3(b) plots the FWHM (black open squares) and height (blue open circles) of a typical Coulomb peak as a function of temperature, as well as the linear fits (dashed lines) of the data. The FWHM of the Coulomb peak increases linearly with the temperature while the peak height remains almost constant.

Three typical Coulomb peaks in the range of $V_{BG}$=12.32–12.43V are also tracked in the same way (labeled as peaks 1, 2, and 3). The FWHM of these peaks as a function of the temperature is shown in Fig. 3(c). The FWHM decreases linearly while lowering the temperature. Using the linear fitting slopes and the previously obtained lever arm, we can estimate the FWHM to be $(1.41\pm0.07)k_BT$.

Using Eq. (1), the FWHM is calculated to be 4.4$k_BT$ in the classical regime, which is almost three times the value measured in the experiment. The difference in slope between the theory and experiment can be caused by many reasons, such as the tunneling rate broadening and larger effective electron temperature.[27] The deviation from the theoretical expectation suggests that the assumption that the effect of $h\Gamma$ can be neglected may not be appropriate. There may also be other reasons which need to be further studied.

Fig. 3(d) shows the Coulomb peak heights of the three peaks as a function of the temperature, exhibiting a behavior that is almost independent of the temperature. This is the signature of Coulomb peaks in the classical regime.[28] Moreover, when changing to the region where a large bias voltage, $V_{SD}$, is applied, another series of Coulomb diamonds with a charging energy $E_C$ larger than 30 meV emerge (Fig. S2 in the supplemental material). Because this series of Coulomb peaks has a much larger $E_C$ and is nearly independent of the gate voltages applied on the top gates, it is considered to be the result of an impurity trap. The effective QD radius of this area is estimated to be 10 nm, resulting in the single particle energy of $\Delta E$=2.6 meV. Compared to $k_BT$~50 μeV, Coulomb oscillations of the impurity trap lies in the quantum regime ($k_BT \ll \Delta E$). Similarly, we investigate the temperature dependence of two typical peaks of the impurity trap (Fig. S3 in the supplemental material). The height of the Coulomb peak decrease linearly with increasing temperature, which is the signature of a Coulomb peak in the quantum resonant tunneling regime.[28]

Fig. 3(e) shows the normalized peak heights as a function of temperature for Coulomb peaks obtained from the QD (red dashed line) and the impurity trap (blue dashed line) on $WS_2$. The peak height of the impurity trap (quantum regime) has a much larger slope that decreases with temperature, which is consistent with theoretical predictions.[28] Meanwhile, the nonzero slope of the QD (classical regime) may be caused by the influence of the impurity in $WS_2$ flake. However, although the impurity trap exists in our device, the behavior of the Coulomb peaks is still in agreement with standard semiconductor QD theory. Note that these results are different from those found with graphene etched QDs, wherein Coulomb peaks belonging to the same QD are able to have different temperature dependences. Among these Coulomb peaks, only portions can be understood by the standard semiconductor theory, with most of the Coulomb peaks becoming broader and higher with increasing temperature.[21,30] The temperature dependence of the normalized peak height of graphene etched QDs is also plotted in Fig. 3(e) for contrast, using a green solid line (The data is from Ref. [21]).

## Discussion

The anomalous behavior of Coulomb peaks in a graphene etched QD is widely considered to be the result of a disordered confining potential[31] since the edge states are formed in the narrow constrictions connecting the QD and the reservoirs. Moreover, this disordered confining potential may limit the performance of the graphene etched QD because low-frequency noise experiments on graphene QDs exhibit a larger fluctuation of potential.[21] Our results, which are consistent with standard semiconductor theory, suggest the existence of an energy-independent tunneling barrier, which is different from the situation existing in graphene QDs. The difference, which relates to the edge states, in the barrier-forming mechanisms of an graphene etched QD and a $WS_2$ QD, may contribute to the larger noise level found in graphene QDs. Meanwhile, the designed device structure used in our experiment, where a disordered confining potential is absent, may be useful for the fabrication of QDs on graphene-like two-dimensional layered materials.

In summary, we use standard semiconductor fabrication techniques to fabricate a QD on WS$_2$, which is one of the TMDCs materials. The FWHM of the Coulomb peaks increases linearly with temperature while the height of the peaks remains nearly independent of temperature, showing that the behavior of the Coulomb oscillations is consistent with standard semiconductor QD theory. Unlike etched graphene QDs, where Coulomb peaks belonging to the same QD can have different temperature dependences, our results indicate the absence of a disordered confining potential. This difference in the potential-forming mechanisms of graphene etched QDs and WS$_2$ QDs may be the reason for the larger potential fluctuation found in graphene QDs. Meanwhile, the designed device structure used in our experiment, which is isolated from the edge states, illustrates the fabrication of QDs on graphene-like two-dimensional layered materials.

**Methods**

The WS$_2$ flakes were produced by mechanically cleaving a bulk WS$_2$ crystal using the "Scotch tape" method onto a highly-doped silicon substrate covered by 100 nm of SiO$_2$, similar to the method used to fabricate graphene devices.[19,30] Few-layer flakes were selected using an optical microscope. The flake of the device studied in the experiment has a thickness of approximately 7-10 layers according to our experience. The source-drain electrodes were formed using a standard electron beam lithography (EBL) process followed by electron beam evaporation to deposit 10 nm of Pd and 90 nm of Au. After a standard lift-off process, an atomic layer deposition (ALD) technique was used to make a 100 nm-thick insulating $Al_2O_3$ layer. Then another EBL step was subsequently applied to form a pattern of four split top gates. Using electron beam evaporation, 5 nm of Ti and 45 nm of Au was deposited to fabricate the top gates. After the $Al_2O_3$ layer covering the source-drain contacts was etched, the device was bonded to the chip carrier and was ready for measurements.

**References**

1   Schwierz, F. Graphene transistors. *Nature Nanotech.* **5**, 487-496 (2010).

2   Geim, A. K. & Novoselov, K. S. The rise of graphene. *Nature Mater.* **6**, 183-191 (2007).

3   Fiori, G. *et al.* Electronics based on two-dimensional materials. *Nature Nanotech.* **9**, 768-779 (2014).

4   Lee, C.-H. *et al.* Atomically thin p-n junctions with van der Waals heterointerfaces. *Nature Nanotech.* **9**, 676-681 (2014).

5   Bolotin, K. I. *et al.* Ultrahigh electron mobility in suspended graphene. *Solid State Commun.* **146**, 351-355 (2008).

6   Wang, Q. H., Kalantar-Zadeh, K., Kis, A., Coleman, J. N. & Strano, M. S. Electronics and optoelectronics of two-dimensional transition metal dichalcogenides. *Nature Nanotech.* **7**, 699-712 (2012).

7   Jariwala, D., Sangwan, V. K., Lauhon, L. J., Marks, T. J. & Hersam, M. C. Emerging Device Applications for Semiconducting Two-Dimensional Transition Metal Dichalcogenides. *ACS Nano* **8(2)**, 1102-1120 (2014).

8   Radisavljevic, B., Radenovic, A., Brivio, J., Giacometti, V. & Kis, A. Single-layer $MoS_2$ transistors. *Nature Nanotech.* **6**, 147-150 (2011).

9   Podzorov, V., Gershenson, M. E., Kloc, C., Zeis, R. & Bucher, E. High-mobility field-effect transistors based on transition metal dichalcogenides. *Appl. Phys. Lett.* **84**, 3301-3303 (2004).

10  Cheng, R. *et al.* Few-layer molybdenum disulfide transistors and circuits for high-speed flexible electronics. *Nature Commun.* **5**, 5143 (2014).

11   Braga, D., Lezama, I. G., Berger, H. & Morpurgo, A. F. Quantitative Determination of the Band Gap of $WS_2$ with Ambipolar Ionic Liquid-Gated Transistors. *Nano Lett.* **12**, 5218-5223 (2012).

12   Bao, W., Cai, X., Kim, D., Sridhara, K. & Fuhrer, M. S. High mobility ambipolar $MoS_2$ field-effect transistors: Substrate and dielectric effects. *Appl. Phys. Lett.* **102**, 042104 (2013).

13   Kang, J., Liu, W. & Banerjee, K. High-performance $MoS_2$ transistors with low-resistance molybdenum contacts. *Appl. Phys. Lett.* **104**, 093106 (2014).

14   Georgiou, T. *et al.* Vertical field-effect transistor based on graphene-$WS_2$ heterostructures for flexible and transparent electronics. *Nature Nanotech.* **8**, 100-103 (2013).

15   Choi, M. S. *et al.* Controlled charge trapping by molybdenum disulphide and graphene in ultrathin heterostructured memory devices. *Nature Commun.* **4**, 1624 (2013).

16   Choi, W. *et al.* High-Detectivity Multilayer $MoS_2$ Phototransistors with Spectral Response from Ultraviolet to Infrared. *Adv. Mater.* **24**, 5832-5836 (2012).

17   Lopez-Sanchez, O., Lembke, D., Kayci, M., Radenovic, A. & Kis, A. Ultrasensitive photodetectors based on monolayer $MoS_2$. *Nature Nanotech.* **8**, 497-501 (2013).

18   Ponomarenko, L. A. *et al.* Chaotic dirac billiard in graphene quantum dots. *Science* **320**, 356-358 (2008).

19   Wang, L.-J. *et al.* A graphene quantum dot with a single electron transistor as an integrated charge sensor. *Appl. Phys. Lett.* **97**, 262113 (2010).

20   Martin, J. *et al.* Observation of electron-hole puddles in graphene using a scanning single-electron transistor. *Nature Phys.* **4**, 144-148 (2008).

21   Song, X.-X. *et al.* Suspending Effect on Low-Frequency Charge Noise in Graphene Quantum Dot. *Sci. Rep.* **5**, 8142 (2015).

22   Kormanyos, A., Zolyomi, V., Drummond, N. D. & Burkard, G. Spin-Orbit Coupling, Quantum Dots, and Qubits in Monolayer Transition Metal

Dichalcogenides. *Phys. Rev. X* **4**, 011034 (2014).

23   Song, X.-X. *et al.* A gate defined quantum dot on the two-dimensional transition metal dichalcogenide semiconductor WSe$_2$. Preprint at http://arxiv.org/abs/1501.04377 (2015).

24   Liu, L., Kumar, S. B., Ouyang, Y. & Guo, J. Performance Limits of Monolayer Transition Metal Dichalcogenide Transistors. *IEEE Trans. Electron Devices* **58**, 3042-3047 (2011).

25   Jo, S., Ubrig, N., Berger, H., Kuzmenko, A. B. & Morpurgo, A. F. Mono- and Bilayer WS$_2$ Light-Emitting Transistors. *Nano Lett.* **14**, 2019-2025 (2014).

26   Ovchinnikov, D., Allain, A., Huang, Y.-S., Dumcenco, D. & Kis, A. Electrical Transport Properties of Single-Layer WS$_2$. *ACS Nano* **8(8)**, 8174-8181 (2014).

27   Kouwenhoven, L. P., Marcus, C. M., Mceuen, P. L., Tarucha, S., Westervelt, R. M. & Wingreen, N. S. Electron transport in quantum dots. *Kluwer Series*, **E345,** Proceedings of the NATO Advanced Study Institute on Mesoscopic Electron Transport, 105–214 (1997).

28   Beenakker, C. W. J. Theory of Coulomb-blockade oscillations in the conductance of a quantum dot. *Phys. Rev. B* **44**, 1646 (1991).

29   Foxman, E. B. *et al.* Effects of quantum levels on transport through a Coulomb island. *Phys. Rev. B* **47**, 10020(R) (1993).

30   Droescher, S., Knowles, H., Meir, Y., Ensslin, K. & Ihn, T. Coulomb gap in graphene nanoribbons. *Phys. Rev. B* **84**, 073405 (2011).

31   Meir, Y., Wingreen, N. S. & Lee, P. A. Transport through a strongly interacting electron system: Theory of periodic conductance oscillations. *Phys. Rev. Lett.* **66**, 3048 (1991)

32   Kumar, A. & Ahluwalia, P. K. Tunable dielectric response of transition metals dichalcogenides MX$_2$ (M=Mo, W; X=S, Se, Te): Effect of quantum confinement. *Physica B* **407**, 4627-4634 (2012).

**Acknowledgments**

This work was supported by the National Fundamental Research Program (Grant No. 2011CBA00200), the National Natural Science Foundation (Grand Nos. 11222438, 11174267, 61306150, 11304301 and 91121014) and the Chinese Academy of Sciences.

**Author contributions**

X.X.S. fabricated the samples. X.X.S., J.Y., D.L. performed the measurements. X.X.S., Z.Z.Z., H.O.L., G.C. and M.X. provided theoretical support. Data are analyzed by X.X.S., J.Y. and Z.Z.Z.. The manuscript is prepared by X.X.S., Z.Z.Z., and G.P.G.. G.P.G. supervised the project. All authors contributed in discussing the results and commented on the manuscript.

**Additional information**

Supplementary information accompanies this paper at ********.
The author(s) declare no competing financial interests.

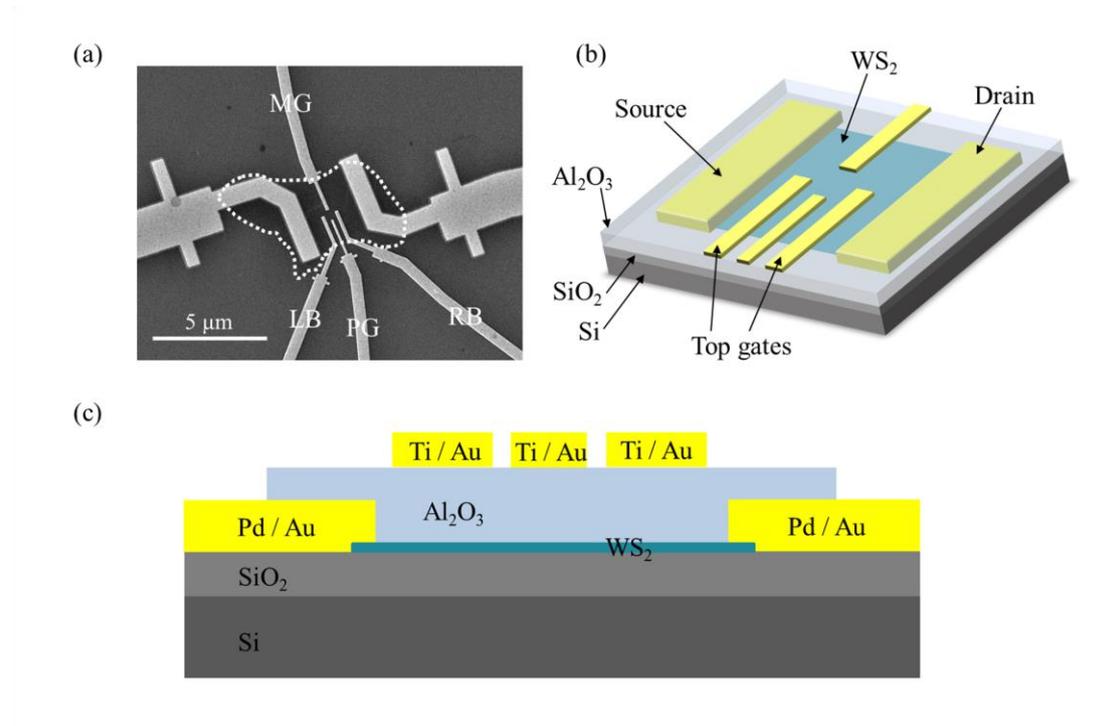

**Figure 1. Device characterization.** (a) Scanning electron microscope image of the WS$_2$ quantum dot studied in this work. The WS$_2$ flake is highlighted by the white dotted line, and the four top gates are labeled as MG, LB, PG, RB. The scale bar represents 5 μm. (b) Three-dimensional schematic view of the device. (c) Schematic cross-section of the device. The few-layer WS$_2$ is deposited on a heavily-doped silicon substrate covered with 100 nm of SiO$_2$. The WS$_2$ flake is separated from the four top gates (Ti/Au) by 100 nm of ALD-grown Al$_2$O$_3$. Two metal gates (Pd/Au) are connected to the flake and used as source-drain contacts.

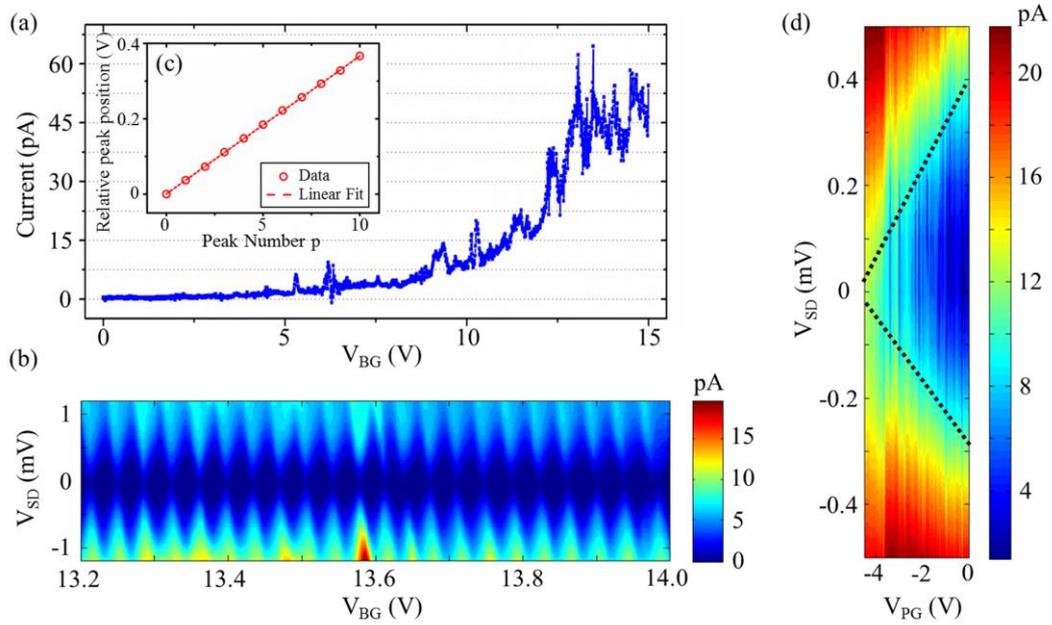

**Figure 2. Transport measurement of the device.** (a) Source-drain current flows through the $WS_2$ devices as a function of back gate voltage, $V_{BG}$, showing the characteristic behavior of an n-doped semiconductor. (b) Over 20 consecutive Coulomb diamonds of a $WS_2$ quantum dot. Symmetric Coulomb diamonds suggests equivalent tunnel coupling to the source and drain leads. All of the top gates have an applied dc voltage of −2 V. (c) The relative peak position as a function of peak number p for the first 10 peaks of (b). The red dashed line is the linear fit for the data (red open circles). (d) A Coulomb diamond measured as a function of the plunger gate voltage $V_{PG}$. Two black dotted lines mark the two sides of the diamond.

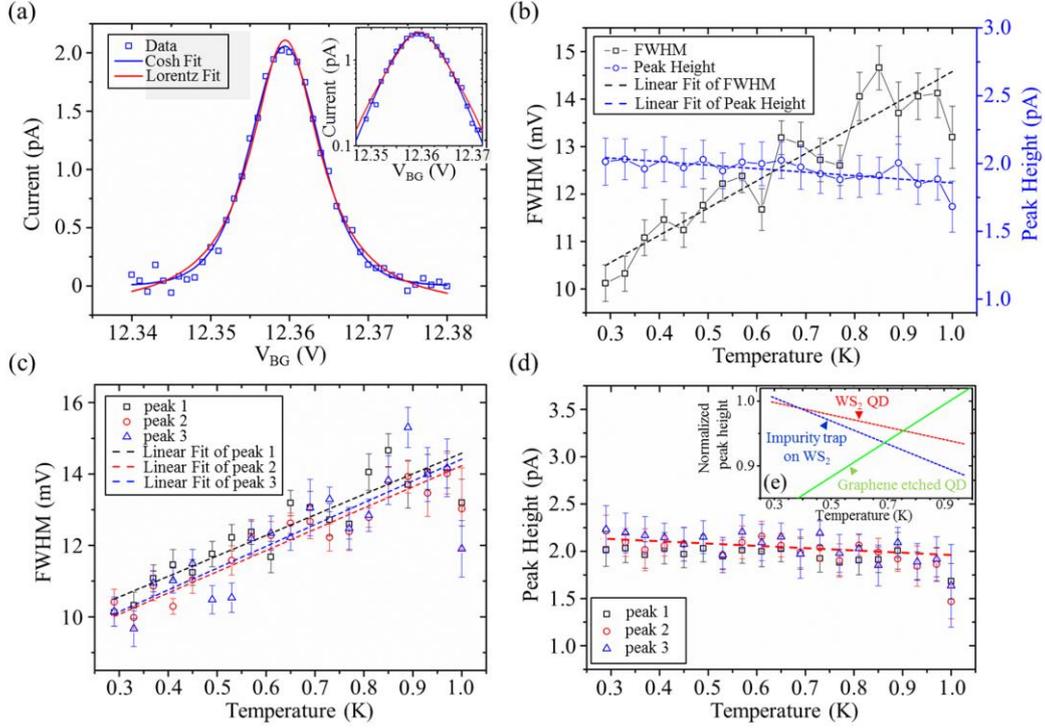

**Figure 3. Temperature dependence of the Coulomb peaks.** (a) A typical Coulomb peak of the QD on $WS_2$ fitted with a Lorentzian (red solid line) and $\cosh(x)$ (blue solid line). (Inset) Same figure plotted in the logarithmic scale. (b) The FWHM (black open squares) and the peak height (blue open circles) of a typical Coulomb peak as a function of temperature. The dashed lines are linear fits. (c) The FWHM of three Coulomb peaks (labeled as peaks 1, 2, and 3) as a function of temperature. The dashed lines show the linear fits for each peak, linearly increasing with temperature. (d) The peak height of the three Coulomb peaks in (c) (labeled as peaks 1, 2, and 3) as a function of temperature. The peak heights are almost independent of temperature. (e) Normalized Coulomb peak height as a function of temperature. The red (blue) dashed line shows the temperature dependence of the peak height obtained from the QD (impurity trap) on $WS_2$, which lies in the classical (quantum) regime. The green solid line indicates the temperature dependence of a common peak height obtained from a graphene etched QD taken from Ref. [21].

# Supplemental Material for:

# Temperature dependence of Coulomb oscillations in a few-layer two-dimensional WS$_2$ quantum dot

Xiang-Xiang Song,[1,2] Zhuo-Zhi Zhang,[1,2] Jie You,[1,2] Di Liu,[1,2] Hai-Ou Li,[1,2] Gang Cao,[1,2] Ming Xiao,[1,2] and Guo-Ping Guo[1,2]

[1] *Key Laboratory of Quantum Information, CAS, University of Science and Technology of China, Hefei, Anhui 230026, China*

[2] *Synergetic Innovation Center of Quantum Information & Quantum Physics, University of Science and Technology of China, Hefei, Anhui 230026, China*

## 1. Estimation of $E_C$ of the WS$_2$ quantum dot

In order to obtain the charging energy $E_C$ precisely, we need to plot the Coulomb diamonds diagram in logarithmic scale. Fig. S1 shows the similar Coulomb diamonds diagram with Fig. 2(a), but in logarithmic scale.

The bias voltage $V_{SD}$ tuned the Fermi level in the source (drain is kept ground), which opens a bias window for the electron to transport through the quantum dot (QD). We can obtain the charging energy $E_C$ from the bias voltage we applied. As shown in Fig. S1, we can estimate that all of the Coulomb diamonds have the same charging energy $E_C$ of approximately 0.75 meV.[S1]

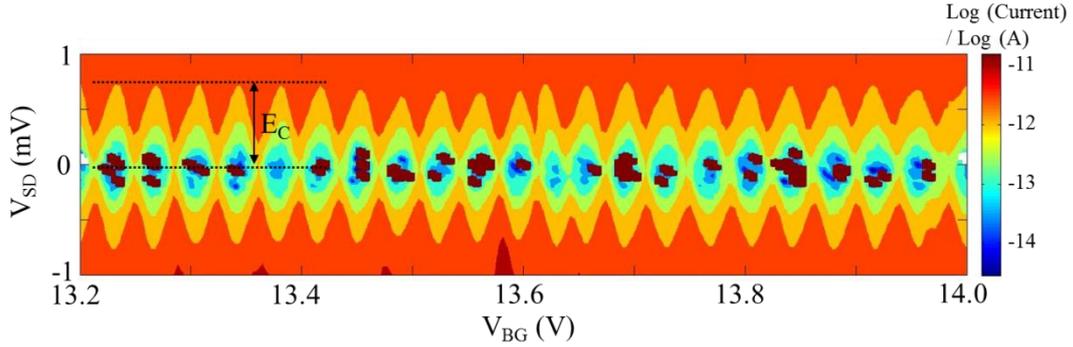

Figure S1. Logarithmic plotting of Coulomb diamonds diagram shown in Fig. 2(a).

## 2. Quantum-dot-like behavior of impurity trap in WS$_2$ device

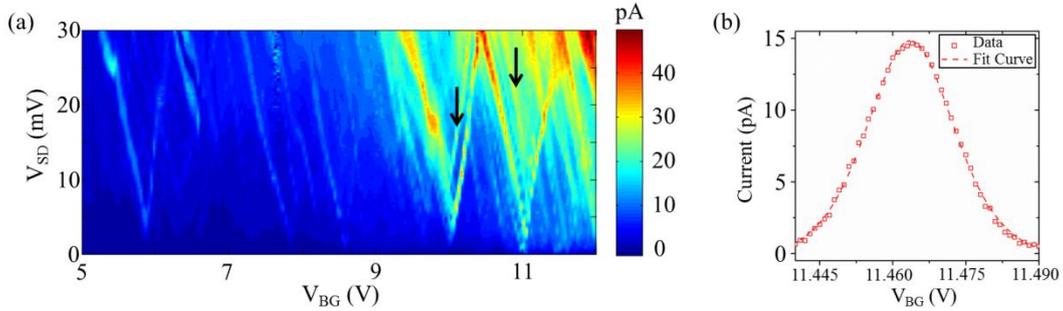

Figure S2. (a) A series of Coulomb diamonds found at larger $V_{SD}$ considered to be the result of an impurity trap. (b) A typical Coulomb peak of the impurity trap fitted with cosh($x$) (red dashed line).

Fig. S2(a) shows the Coulomb diamonds found at larger $V_{SD}$ values. This series of Coulomb diamonds has a charging energy $E_C$ larger than 30 meV and exhibits more than one period. The effective "dot" radius is estimated to be 10 nm. In addition, this series of Coulomb peaks is almost independent of the voltages applied to the top gates. All of this is evidence of an impurity trap that behaves like a quantum dot (QD).[S2] Fig. S2(b) shows a typical Coulomb peak of the impurity trap. Because the "QD" radius is 10 nm, a much larger single particle energy, $\Delta E$=2.6 meV, is obtained. This value is consistent with the experiment because evidence of the presence of an excited state is observed, indicated by the black arrows in the Coulomb diamond diagram (Fig. S2(a)). It is estimated that $k_BT\sim 50$ μeV, suggesting that the Coulomb peaks of the impurity trap are in the quantum regime ($k_BT \ll \Delta E$). A cosh(*x*) fit is used to fit the Coulomb peak, given by the red dashed line in Fig. S2(b).

## 3. Temperature dependence of impurity trap Coulomb peaks

Similar to the $WS_2$ QD, we investigate the temperature dependence of two typical peaks of the impurity trap (labeled peaks 4 and 5). To exclude the influence of the QD, a bias voltage of $V_{SD}\sim 1$ mV is applied.

Fig. S3(a) shows the peak height as a function of temperature. The height of the Coulomb peak decreases linearly with increasing temperature, which is the signature of a Coulomb peak in the quantum resonant tunneling regime.[S3] The temperature dependence of the full-width-at-half-maximum (FWHM) of the Coulomb peak is shown in Fig. S3(b). Different from the predicted linear dependence,[S3] the FWHM remains almost constant from 290 mK to 1 K. This also can be explained by the standard semiconductor QD theory.[S1] Because it is necessary to avoid the influence of the QD when detecting the current of the impurity trap, a typical dc bias of 1 mV is applied. This voltage $V_{SD}$ gives an energy of approximately 1 meV to the "dot", which corresponds to an energy of 11.6 K. Under these circumstances, the temperature can no longer dominate the FWHM of the Coulomb peak. Such bias voltage also contributes to the different range of the Coulomb peaks obtained from the QD and the impurity trap.

Although the "QD" is formed owing to the impurity trap, the behavior of the Coulomb peaks is also in agreement with standard semiconductor QD theory.

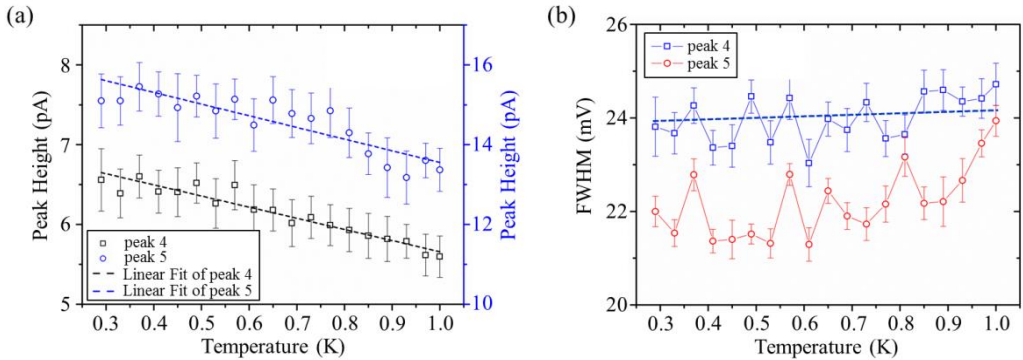

Figure S3. (a) The peak height of two typical Coulomb peaks (labeled peaks 4 and 5) of the impurity trap as a function of temperature, including the linear fits (dashed lines), exhibiting a linear decrease with temperature. (b) Temperature dependence of the FWHM of the two Coulomb peaks in (a) (labeled peaks 4 and 5).

# REFERENCES

1. L. P. Kouwenhoven, C. M. Marcus, P. L. Mceuen, S. Tarucha, R. M. Westervelt, and N. S. Wingreen, Electron transport in quantum dots. *Kluwer Series*, **E345,** Proceedings of the NATO Advanced Study Institute on Mesoscopic Electron Transport, 105-214 (1997).
2. M. Sanquer, M. Specht, L. Ghenim, S. Deleonibus, and G. Guegan, Coulomb blockade in low-mobility nanometer size Si MOSFET's. *Phys. Rev. B* **61,** 7249 (2000).
3. C. W. J. Beenakker. Theory of Coulomb-blockade oscillations in the conductance of a quantum dot. *Phys. Rev. B* **44,** 1646 (1991).